\documentclass[reprint,amsmath,amssymb,nofootinbib,aps,prd]{revtex4-1}

\usepackage{graphicx}
\usepackage{dcolumn}
\usepackage{bm}

\usepackage{hhline}
\usepackage[american]{babel}
\usepackage{booktabs}
\usepackage{dcolumn}  
\usepackage[T1]{fontenc}  
\usepackage[colorlinks]{hyperref}  
\usepackage[utf8]{inputenc}  
\usepackage{multirow}
\usepackage{times}  
\usepackage{url}
\usepackage[dvipsnames]{xcolor}
\usepackage{float}
\usepackage{resizegather}

\newcolumntype{d}[1]{D{.}{.}{#1}}
\newcolumntype{v}[1]{D{,}{,\ }{#1}}

\begin{document}
	
	
	\title{Spherically symmetric solutions in higher-derivative theories of	gravity}
	
	\author{G. Rodrigues-da-Silva}
	\email{gesiel.fisica@gmail.com}
	\affiliation{Departamento de Física, Universidade Federal do Rio Grande do Norte,\\
	Campus Universitário, s/n - Lagoa Nova, CEP 59072-970, Natal, Rio Grande do Norte, Brazil}
	
	\author{L. G. Medeiros}
	\email{leogmedeiros@ect.ufrn.br}
	\affiliation{Escola de Ciências e Tecnologia, 	Universidade Federal do Rio Grande do Norte,\\
	Campus Universitário, s/n - Lagoa Nova, CEP 59072-970, Natal, Rio Grande do Norte, Brazil}

	\date{\today}

\begin{abstract}

Higher-order theories of gravity have received much attention from several
areas including quantum gravity, string theory and cosmology. This paper
proposes a higher-order gravity whose action includes all curvature scalar terms up to the second-order corrections of general relativity, namely,
$R^{2}$, $R^{3}$ and $R\square R$\emph{.} Then, we explore spherically
symmetric and static solutions in the weak-field regime and black holes
context. All solutions in the weak-field regime due to a point mass are
deduced, and by making a stability analysis of these solutions, we restrict
them to Yukawa-type solutions. In regard to black hole solutions, we
use the Lichnerovicz method to investigate the possibility of existence of
non-Schwarzschild black holes. The results obtained show that
non-Schwarzschild solutions might exist. However, for reasonable values of the
parameters of the theory, its horizon radii are extremely small making
macroscopic black holes different from Schwarzschild unfeasible.

\end{abstract}

\maketitle

\section{\label{Introduction} Introduction}

General relativity (GR) has still been our standard theory for describing
gravity. This is due to its immense predictive power, which even after a
century it continues to be corroborated by observations, namely, direct
detections of gravitational waves from the binary black hole and neutron star
mergers \cite{Abbott:2016blz,TheLIGOScientific:2016src,PhysRevLett.119.161101}.

Despite the great success of GR, extensions to it have received much attention
from several areas including high-energy physics, e.g. string theory, the
cosmology of the early and late universe, and astrophysics.

From a theoretical point of view, one of the limitations of GR is the fact
that it is non-renormalizable, making its quantization problematic. In this
sense, a very relevant kind of extended theory of gravity is the
higher-order one, in which the Einstein-Hilbert action is supplemented by higher-order curvature terms. From this perspective, GR is viewed as an effective low-energy theory that requires higher-order corrections as we
increase the energy scale \cite{AIHPA_1974__20_1_69_0}. Its relevance is
justified since in order to construct a quantum theory of gravity it was found
that higher-order terms contribute to the renormalizability of the theory
\cite{doi.org/10.1063/1.1724264}. In fact, K. S. Stelle showed that by
adding to Einstein-Hilbert action all relevant curvature terms up to the
second-order, one obtains a perturbatively renormalizable system, however, at
the price of introducing ghost-type instabilities, which present themselves as
states with negative energy or negative norm \cite{PhysRevD.16.953}.
Classically, as presented by Ostrogradski, due to the presence of higher-order
derivatives, in general, the Hamiltonian of the system becomes not bounded
from below \cite{2015arXiv150602210W}. This is the known tension between
renormalizability and unitarity \cite{Asorey:1996hz,Shapiro:2008sf,article}. Allying one to another represents one of the biggest
problems in quantum gravity, and that's one of the reasons why we still have
no fully consistent model capable of describing gravity in the ultraviolet
(UV) regime.

In order to construct a quantum theory of gravity, addressing the
renormalizability problem, we highlight two interesting approaches to
extensions of GR. One of those is based on gravities constructed by the
Einstein-Hilbert action supplemented by terms involving a polynomial function
of the covariant d'Alembertian operator, such as $RF_{1}\left(  \square
\right)  R$ and $R_{\mu\nu}F_{2}\left(  \square\right)  R^{\mu\nu}$
\cite{Asorey:1996hz,Modesto:2016ofr}. A theory in which corrections of
this type are included has a finite number of curvature derivatives and is
(super-)renormalizable and local. It is also worth mentioning that in this
approach the ghosts problem can be addressed through the quantization process
\textit{à la} Lee-Wick \cite{PhysRevD.2.1033,Donoghue:2019fcb}. The other
approach involves replacing polynomial with non-polynomial operator functions
of the covariant d'Alembertian, such as $R\exp\left(  \square\right)  R$ and
$R_{\mu\nu}\exp\left(  \square\right)  R^{\mu\nu}$, that makes a theory with
infinite curvature derivatives, (super-)renormalizable, non-local and
ghost-free at the tree-level \cite{Tomboulis:1997gg,Modesto:2011kw,Biswas:2011ar,Shapiro:2015uxa}.

Although we cite renormalizability and unitarity as strong theoretical
motivations driving the search for higher-order theories of gravity, we are
not interested in proposing a model in the quantum gravity scenario. Instead,
our interest lies in classical models of gravity motivated by the context of
effective theory, in which the included higher-order curvature terms become
more relevant in a given energy scale. Some examples of works that consider
higher-order gravities including curvature derivatives terms are: Refs.
\cite{Quandt:1990gc,Accioly:2016qeb,Giacchini:2018gxp}, in which
the authors analyze low-energy effects; Refs. \cite{Cuzinatto:2006hb,Cuzinatto:2016ehv}, in which they develope the gauge formulation and a
study of equivalence between theories, respectively; and Refs.
\cite{Castellanos:2018dub,Cuzinatto:2018vjt,Cuzinatto:2018chu}, in
which the authors propose an extension to the Starobinsky's inflationary model.

For the sake of simplicity, the gravity treated here is built only with the
curvature scalar $R$ and its derivatives --- we will not include, for example,
terms like $R_{\mu\nu}R^{\mu\nu}$ or $R_{\mu\nu}F_{2}\left(  \square\right)
R^{\mu\nu}$. It is characterized by including all correction terms up to the
second-order, that is, quadratic and cubic curvature correction terms involving the curvature scalar, i.e. $R^{2}$, $R^{3}$ and $R\square R$. This implies sixth-order field equations for
the metric. Thus, the most general gravitational action up to the second-order and
containing only the curvature scalar is given by%
\begin{gather}
	S=\frac{M_{P}^{2}}{2}\int d^{4}x\sqrt{-g}\left(  R+\frac{1}{2\kappa_{0}}%
	R^{2}+\frac{\alpha_{0}}{3\kappa_{0}^{2}}R^{3}-\frac{\beta_{0}}{2\kappa_{0}%
		^{2}}R\square R\right) + \nonumber\\  +\int d^{4}x\sqrt{-g}%
	\mathcal{L}%
	_{m}, 
	\label{eq:acao-da-teoria}
\end{gather}
in which $M_{P}$ is the reduced Planck mass, so that $M_{P}^{2}\equiv\left(
8\pi G\right)  ^{-1}$, and $\kappa_{0}$, $\alpha_{0}$ and $\beta_{0}$ are
parameters of the theory, $\kappa_{0}$ being the only dimensional one (squared
mass). The quantity $%
\mathcal{L}%
_{m}=%
\mathcal{L}%
_{m}\left(  \psi,g_{\mu\nu}\right)  $ represents the matter Lagrangian, which
encapsulates the coupling of gravitation with the matter fields $\psi $.

One of the fundamental aspects in which the model defined by Eq.
(\ref{eq:acao-da-teoria}) can be addressed is in the study and analysis of its
solutions. In the late 1970s, spherically symmetric solutions of the gravity
with all quadratic curvature corrections were studied by Stelle
\cite{Stelle:1977ry}. In that paper, the weak-field regime of the theory became
well established: each of the massive modes contributes to a decreasing
Yukawa-type term, while the usual non-massive spin-2 mode contributes to the
Newtonian potential. In addition, a full analysis was also performed by assuming a
spherically symmetric and static metric in Schwarzschild coordinates, and by
taking an expansion in Frobenius series around the origin of the two radial
functions of the metric. However, as justified in Ref. \cite{PhysRevD.92.124019},
at that pre computational algebra time, it was not possible to have a complete
view of the picture. This subject has recently been revived when, using
Lichnerovicz and Israel type no-hair theorems, assuming the existence of a
horizon $r_{h}$, along with asymptotic flatness at infinity, William Nelson
has shown that not only the curvature scalar $R$ but also the Ricci tensor $R_{\mu\nu
}$ would disappear outside the horizon \cite{Nelson:2010ig}. This would mean
that the Schwarzschild solution was the only spherically symmetric static
solution of the analyzed gravity. On the other hand, it was shown in Ref. \cite{Lu:2015cqa} that signal errors were made in the analysis of the
traceless part, so that the Ricci tensor $R_{\mu\nu}$ would not disappear in
the exterior region. This was an indication that non-Schwarzschild solutions
might exist. In fact, families of non-Schwarzschild solutions were found
numerically by Ref. \cite{Lu:2015cqa}, and then other papers confirmed these
results using analytical \cite{Podolsky:2018pfe,Podolsky:2019gro} and semi-analytical
\cite{Kokkotas:2017zwt} approaches.

In this paper, we explore spherically symmetric and static solutions in the context of the weak-field regime and black hole solutions. Initially in Section
\ref{m:weak-field-limit}, we present the field equations of the gravity
(\ref{eq:acao-da-teoria}) and then we deduce all solutions in the weak-field
regime due to a point mass. Similar to the analysis developed in Ref.
\cite{Quandt:1990gc}, these solutions are obtained using an ansatz for the
curvature scalar $R$ and for the gravitational potential $\phi$, and by
establishing the boundary conditions, which are, a point mass at the origin
and asymptotic flatness at the infinite. At first, among the solutions found are the decreasing Yukawa-type, oscillatory solutions and
hybrid ones, involving Yukawa-type and oscillatory terms. However, in
Subsection \ref{m:stability}, a stability analysis of these solutions
restricts the range of the theory parameters, discarding all solutions
involving oscillatory terms. Regarding black hole solutions, in Section
\ref{m:black-hole-solutions}, we use the Lichnerovicz method, similar to the
found in Refs. \cite{Nelson:2010ig,PhysRevD.92.124019}, to investigate
the possibility of existence of exterior non-Schwarzschild solutions. At first, we use this method considering
the full equations of the theory and, based on the Planck and inflationary
energy scales, we estimate the order of magnitude of the horizon for the
occurrence of such a non-Schwarzschild black hole. Then, in the Subsection
\ref{m:deviations}, we explore the existence of deviations from the exterior
Schwarzschild solution, and for that, we assume a continuous deformation of
the curvature scalar. With this analysis we confirm in an alternative way the
previous results of the Section \ref{m:black-hole-solutions}. Finally, our
conclusions are presented in Section \ref{m:discussions}.

All over the work, the chosen metric signature is $\left(  -,+,+,+\right)
$,\emph{ }the Riemann tensor is written as $R^{\kappa}{}_{\lambda\mu\nu}%
\equiv\partial_{\mu}\Gamma_{\lambda\nu}^{\kappa}-\partial_{\nu}\Gamma
_{\lambda\mu}^{\kappa}+\Gamma_{\rho\mu}^{\kappa}\Gamma_{\lambda\nu}^{\rho
}-\Gamma_{\rho\nu}^{\kappa}\Gamma_{\lambda\mu}^{\rho}$ and the Ricci tensor is
defined as $R_{\lambda\nu}\equiv R^{\mu}{}_{\lambda\mu\nu}$.

\section{Field equations and weak-field solutions \label{m:weak-field-limit}}

The field equations in the metric formalism associated with the action
(\ref{eq:acao-da-teoria}) can be written as%

\begin{widetext}

\begin{gather}
	R_{\mu\nu}-\frac{1}{2}g_{\mu\nu}R+\frac{1}{\kappa_{0}}\left[
	R\left(  R_{\mu\nu}-\frac{1}{4}g_{\mu\nu}R\right)  -\nabla_{\mu}\nabla_{\nu
	}R+g_{\mu\nu}\square R\right] 
	+\frac{\alpha_{0}}{\kappa_{0}^{2}}\left[  R^{2}\left(  R_{\mu\nu}-\frac{1}%
	{6}g_{\mu\nu}R\right)  -\nabla_{\mu}\nabla_{\nu}R^{2}+g_{\mu\nu}\square
	R^{2}\right]  +\nonumber\\
	+\frac{\beta_{0}}{\kappa_{0}^{2}}\left[  \nabla_{\mu}\nabla_{\nu}\square
	R+\frac{1}{2}\nabla_{\mu}R\nabla_{\nu}R-R_{\mu\nu}\square R-g_{\mu\nu}\left(
	\square^{2}R+\frac{1}{4}\nabla_{\alpha}R\nabla^{\alpha}R\right)  \right]
	=8\pi GT_{\mu\nu}, \label{eq:eqs-de-campo}%
\end{gather}

\end{widetext}
	
where\emph{ }$\square\equiv\nabla^{\mu}\nabla_{\mu}$\emph{ }and%
\begin{equation}
T_{\mu\nu}\equiv\frac{2}{\sqrt{-g}}\frac{\delta\left(  \sqrt{-g}%
	\mathcal{L}%
	_{m}\right)  }{\delta g^{\mu\nu}}.
\end{equation}
These are sixth-order field equations for the metric. As already said, the
terms beyond Einstein-Hilbert in the field equations (\ref{eq:eqs-de-campo})
become increasingly relevant in the UV regime and they are regulated by powers
of $\kappa_{0}$ in the denominator.

From this point forward until the end of the section, we focus on the
weak-field limit analysis of the gravity (\ref{eq:acao-da-teoria}). By
following the approach developed in Ref. \cite{Quandt:1990gc},\emph{\ }we analyze in detail all possible solutions in the weak-field regime due to the
higher-order corrections.

In the weak-field approximation, in which $g_{\mu\nu}=\eta_{\mu\nu}+h_{\mu\nu
}$, with $\left\vert h_{\mu\nu}\right\vert \ll1$, the field equations can be
expressed as%

\begin{gather}
	R_{\mu\nu}^{lin}-\frac{1}{2}\eta_{\mu\nu}R^{lin}+\frac{1}{\kappa_{0}}\left(
	\eta_{\mu\nu}\square R^{lin}-\partial_{\mu}\partial_{\nu}R^{lin}\right)
	+\nonumber\\
	+\frac{\beta_{0}}{\kappa_{0}^{2}}\left(  \partial_{\mu}\partial_{\nu}\square
	R^{lin}-\eta_{\mu\nu}^{2}\square R^{lin}\right)  =8\pi GT_{\mu\nu
	},\label{eq:eqs-campo-fraco}%
\end{gather}
since we ignore all nonlinear terms in $h_{\mu\nu}$. In Eq.
(\ref{eq:eqs-campo-fraco}), the superscript \textit{lin} indicates that the
quantities are linearized. Moreover, in this regime the box operator
simplifies to $\square=\partial^{\mu}\partial_{\mu}$. By taking its trace and
considering an energy-momentum tensor $T^{\mu\nu}$ of a dust-like fluid,
namely, $T^{\mu\nu}=\rho u^{\mu}u^{\nu}$, where $u^{0}\gg u^{i}$, we get%
\begin{equation}
R^{lin}-\frac{3}{\kappa_{0}}\left(  1-\frac{\beta_{0}}{\kappa_{0}}%
\square\right)  \square R^{lin}=8\pi G\rho,\label{eq:eq-traco-com-tempo}%
\end{equation}
which in a static regime reduces to%
\begin{equation}
R^{lin}-\frac{3}{\kappa_{0}}\left(  1-\frac{\beta_{0}}{\kappa_{0}}\nabla
^{2}\right)  \nabla^{2}R^{lin}=8\pi G\rho.\label{eq:eq-traco}%
\end{equation}
On the other hand, by relating Eq. (\ref{eq:eq-traco}) to the component
$\mu=\nu=0$ of Eq. (\ref{eq:eqs-campo-fraco}), in which the geodesic equation
gives us $h_{00}=-2\phi$, with $\phi$ representing the classical gravitational
potential, we obtain
\begin{equation}
\nabla^{2}\phi+\frac{1}{6}R^{lin}=\frac{16}{3}\pi G\rho
.\label{eq:escalar-curvatura-limite}%
\end{equation}
The expression (\ref{eq:eq-traco}) represents a fourth-order differential
equation for the curvature scalar $R^{lin}$. Thus, considering a spherically
symmetric distribution, in which $\rho =M\delta (r)$, it has four linearly
independent solutions. However, if we impose the condition of asymptotic
flatness at infinity, in which $R\rightarrow0$, we reduce the number of
solutions for only two. Similarly,\emph{ }the general solution of Eq.
(\ref{eq:escalar-curvatura-limite}) consists of six linearly independent
solutions, and the physical imposition of asymptotic flatness at infinity
discards three of them.

In order to obtain the solutions of Eq. (\ref{eq:eq-traco}), we propose the
ansatz%
\begin{equation}
R^{lin}=\frac{b_{+}}{r}e^{-\frac{r}{l_{+}}}+\frac{b_{-}}{r}e^{-\frac{r}{l_{-}%
}},\label{eq:proposta-curvatura-escalar}%
\end{equation}
where $b_{+}$,\ $b_{-}$, $l_{+}$\ and $l_{-}$ are constants to be determined
and $\operatorname{Re}\left(  l_{\pm}\right)  $ $\geq0$. Moreover, by
considering a point mass $M$ at the origin, so that $\rho =M\delta (r)$,
and using%
\begin{equation}
\nabla^{2}\left(  \frac{1}{r}e^{-\frac{r}{l}}\right)  =\frac{1}{rl^{2}%
}e^{-\frac{r}{l}}-4\pi\delta\left(  r\right)  ,
\end{equation}
with%
\begin{equation}
\nabla^{2}\left(  \frac{1}{r}\right)  =-4\pi\delta\left(  r\right)  ,
\end{equation}
we obtain for Eq. (\ref{eq:eq-traco}) the expression
\begin{gather}
	\frac{b_{+}}{r}e^{-\frac{r}{l_{+}}}\left(  1-\frac{3}{\kappa_{0}}\frac
	{1}{l_{+}^{2}}+\frac{3\beta_{0}}{\kappa_{0}^{2}}\frac{1}{l_{+}^{4}}\right)
	+\nonumber\\
	+\frac{b_{-}}{r}e^{-\frac{r}{l_{-}}}\left(  1-\frac{3}{\kappa_{0}}\frac
	{1}{l_{-}^{2}}+\frac{3\beta_{0}}{\kappa_{0}^{2}}\frac{1}{l_{-}^{4}}\right)
	+\nonumber\\
	+\frac{12}{\kappa_{0}}\left[  b_{+}+b_{-}-\frac{\beta_{0}}{\kappa_{0}}\left(
	\frac{b_{+}}{l_{+}^{2}}+\frac{b_{-}}{l_{-}^{2}}\right)  \right]  \pi
	\delta\left(  r\right)  +\nonumber\\
	-\frac{12\beta_{0}}{\kappa_{0}^{2}}\left(  b_{+}+b_{-}\right)  \pi\nabla
	^{2}\delta\left(  r\right)  =8\pi GM\delta\left(  r\right) .
\end{gather}
Due to the independence of the various terms in the previous expression, we
then get the following relations:%
\begin{equation}
b_{+}+b_{-}=0,
\end{equation}%
\begin{equation}
\frac{3}{\kappa_{0}}\left[  b_{+}+b_{-}-\frac{\beta_{0}}{\kappa_{0}}\left(
\frac{b_{+}}{l_{+}^{2}}+\frac{b_{-}}{l_{-}^{2}}\right)  \right]
=2GM,\label{eq:seg-relacao2}%
\end{equation}
and%
\begin{equation}
l_{\pm}^{4}-\frac{3}{\kappa_{0}}l_{\pm}^{2}+\frac{3\beta_{0}}{\kappa_{0}^{2}%
}=0.\label{eq:biquadratica2}%
\end{equation}
The coefficientes $l_{\pm}$ can be obtained by solving the biquadratic
equation (\ref{eq:biquadratica2}). Thus%
\begin{equation}
l_{\pm}=\sqrt{\frac{3}{2\kappa_{0}}\left(  1\pm\sqrt{1-\frac{4\beta_{0}}{3}%
	}\right)  },\label{eq:raiz-li}%
\end{equation}
and since $b_{+}+b_{-}=0$, we can get from Eq. (\ref{eq:seg-relacao2}) that%
\begin{equation}
b_{+}=\frac{2GM\kappa_{0}}{3}\frac{1}{\sqrt{1-\frac{4\beta_{0}}{3}}}%
=-b_{-}.\label{eq:b0}%
\end{equation}
In the limit $\left\vert \beta_{0}\right\vert \rightarrow0$, which represents
the weak-field limit of $R^{2}$ gravity, we can write%
\begin{equation}
l_{+}=\sqrt{\frac{3}{\kappa_{0}}}\text{ \ and\ \ }l_{-}=0.
\end{equation}
Thus, for $\kappa_{0}>0$ the $l_{+}$ is associated with Yukawa term due to
$R^{2}$ correction and $l_{-}$ makes the other exponential in the
expression (\ref{eq:proposta-curvatura-escalar}) negligible.

In order to obtain the solution of Eq. (\ref{eq:escalar-curvatura-limite}), we
use again that $ \rho =M \delta (r)$, the Eq.
(\ref{eq:proposta-curvatura-escalar}) for the curvature scalar $R^{lin}$ and
the ansatz%
\begin{equation}
\phi=-\frac{GM}{r}\left(  1+a_{+}e^{-\frac{r}{l_{+}}}+a_{-}e^{-\frac{r}{l_{-}%
}}\right)  , \label{eq:proposta-potencial-gravitacional}%
\end{equation}
in which $a_{+}$\ and $a_{-}$ are constants to be determined.

By following the same procedure performed previously, we obtain%

\begin{equation}
a_{+}=\frac{1}{6}\left(  \frac{1+\sqrt{1-\frac{4\beta_{0}}{3}}}{\sqrt
	{1-\frac{4\beta_{0}}{3}}}\right)  \text{ \ and\ \ }a_{-}=-\frac{1}{6}\left(
\frac{1-\sqrt{1-\frac{4\beta_{0}}{3}}}{\sqrt{1-\frac{4\beta_{0}}{3}}}\right)
.\label{eq:coeficientes-ai}%
\end{equation}
Thus, the generalization for the Newtonian potential is given by Eq.
(\ref{eq:proposta-potencial-gravitacional}) with coefficients
(\ref{eq:raiz-li}) and (\ref{eq:coeficientes-ai}). It is interesting to
evaluate the gravitational potential $\phi$ at the limit where $\beta
_{0}\rightarrow0$. In this case, we have%
\begin{equation}
a_{+}=\frac{1}{3}\text{, \ \ }a_{-}=0\text{ \ and\ }l_{+}=\sqrt{\frac
	{3}{\kappa_{0}}},
\end{equation}
so that%
\begin{equation}
\phi=-\frac{GM}{r}\left(  1+\frac{1}{3}e^{-\sqrt{\frac{\kappa_{0}}{3}}%
	r}\right)  .\label{eq:campo-fraco-limite-b0}%
\end{equation}

In order to investigate the nature of the all possible solutions, we perform a
case analysis of the relations (\ref{eq:raiz-li}).\emph{ }By defining the
\textbf{real positive}\emph{ }quantities%
\begin{gather*}
	A_{\pm}=1\pm\sqrt{1-\frac{4\beta_{0}}{3}}\text{ \ with\emph{ \ }}0<\beta
	_{0}<3/4,\\
	B_{\pm}=\sqrt{1+\frac{4\left\vert \beta_{0}\right\vert }{3}}\pm1\text{
		\emph{\ }with\emph{ }\ }\beta_{0}<0,\\
	C_{\pm}=\sqrt{\sqrt{\frac{4\beta_{0}}{3}}\pm1}\text{ \emph{\ }with\emph{ }%
		\ }\beta_{0}>3/4,
\end{gather*}
all possible solutions can be summarized in Tables \ref{tab:table1} and \ref{tab:table2}:
\begin{table}[h]
\caption{\label{tab:table1}This table shows the values obtained for $l_{\pm}$ in each of the intervals of $\beta_{0}$ for positive $\kappa_{0}$. Note that only the range $0<\beta_{0}<3/4$ gives us real quantities for both $l_{\pm}$.}
\begin{tabular}
[c]{|c|c|c|}\hline
\multicolumn{3}{|c|}{$\kappa_{0}>0$}\\\hline \hline
$\beta_{0}$ & $l_{+}$ & $l_{-}$\\\hline
$0<\beta_{0}<3/4$ & $\sqrt{\frac{3}{2\kappa_{0}}A_{+}}$ & $\sqrt{\frac
	{3}{2\kappa_{0}}A_{-}}$\\\hline 
$\beta_{0}<0$ & $\sqrt{\frac{3}{2\kappa_{0}}B_{+}}$ & $i\sqrt{\frac{3}%
	{2\kappa_{0}}B_{-}}$\\\hline
$\beta_{0}>3/4$ & $\sqrt{\frac{3}{4\kappa_{0}}}\left(  C_{+}+iC_{-}\right)  $
& $\sqrt{\frac{3}{4\kappa_{0}}}\left(  C_{+}-iC_{-}\right)  $ \\\hline
\end{tabular}
\end{table}

\begin{table}[h]
\caption{\label{tab:table2}This table shows the values obtained for $l_{\pm}$ in each of the intervals of $\beta_{0}$ for negative $\kappa_{0}$. None of them gives real values for both $l_{\pm}$.}
\begin{tabular}
[c]{|c|c|c|}\hline
\multicolumn{3}{|c|}{$\kappa_{0}<0$}\\\hline \hline
$\beta_{0}$ & $l_{+}$ & $l_{-}$\\\hline
$0<\beta_{0}<3/4$ & $i\sqrt{\frac{3}{2\left\vert \kappa_{0}\right\vert }A_{+}%
}$ & $i\sqrt{\frac{3}{2\left\vert \kappa_{0}\right\vert }A_{-}}$\\\hline
$\beta_{0}<0$ & $i\sqrt{\frac{3}{2\left\vert \kappa_{0}\right\vert }B_{+}}$ &
$\sqrt{\frac{3}{2\left\vert \kappa_{0}\right\vert }B_{-}}$\\\hline
$\beta_{0}>3/4$ & $\sqrt{\frac{3}{4\left\vert \kappa_{0}\right\vert }}\left(
C_{-}-iC_{+}\right)  $ & $\sqrt{\frac{3}{4\left\vert \kappa_{0}\right\vert }%
}\left(  C_{-}+iC_{+}\right)  $\\\hline
\end{tabular}
\end{table}

The $\kappa_{0}>0$ with $0<\beta_{0}<3/4$ is the only case where $l_{\pm}\in%
\mathbb{R}
$, implying Yukawa corrections $r^{-1}e^{-r/l_{\pm}}$ to the gravitational
potential. On the other hand, $\kappa_{0}<0$ with $0<\beta_{0}<3/4$ is the
only case where $l_{\pm}$ are pure imaginary, implying oscillatory
corrections. In all other cases, $l_{\pm}\in%
\mathbb{C}
$, and the corrections contain Yukawa-type parts and oscillating ones.

It is interesting to note that the real parts $\operatorname{Re}\left(
l_{\pm}\right)  $ are always nonnegative. Thus, all the presented cases
represent good solutions for $R$, since all of them are compatible with the
established boundary conditions at the infinite. In some cases, as we can
typically see when $\kappa_{0}>0$ and $\beta_{0}<0$, one solution has an
oscillatory behavior, however, the factor $r$ in the denominator of
(\ref{eq:proposta-curvatura-escalar}) causes $R\rightarrow0$ at infinity.
Nevertheless, a relevant question is: are all the shown cases above physically
feasible, or are there any intervals at which the parameters $\kappa_{0}$ and
$\beta_{0}$ must be restricted? By investigating the stability of solutions to
gravitational potential $\phi$, we will answer this question.

\subsection{Stability analysis \label{m:stability}}

In order to restrict the possible values that the parameters $\kappa_{0}$ and
$\beta_{0}$ may assume, we analyze the stability of the solutions in the
weak-field limit, similar to that developed in Ref.
\cite{Perivolaropoulos:2019vkb}.

As we can see in Eq. (\ref{eq:escalar-curvatura-limite}) the gravitational
potential $\phi$ depends on the curvature scalar $R$, such that, any
instability in $R$ produces an instability in $\phi$. So, the stability
analysis can be performed by perturbing Eq. (\ref{eq:eq-traco-com-tempo})
assuming that%
\begin{equation}
R=R_{0}\left(  r\right)  +\delta R\left(  r,t\right)  .
\end{equation}
This perturbation must in principle be induced by a perturbation in $\rho$
such that $\rho\rightarrow\rho_{0}+\delta\rho$. In this case, we have for the
perturbed equation%
\begin{equation}
\delta R-\frac{3}{\kappa_{0}}\left[  1-\frac{\beta_{0}}{\kappa_{0}}\left(
-\partial_{0}^{2}+\nabla^{2}\right)  \right]  \left(  -\partial_{0}^{2}%
+\nabla^{2}\right)  \delta R=8\pi G\delta\rho. \label{Eq4}%
\end{equation}
By taking the Fourier transform of Eq. (\ref{Eq4}), we have%
\begin{equation}
\delta R_{k}+\frac{3}{\kappa_{0}}\left[  1+\frac{\beta_{0}}{\kappa_{0}}\left(
\partial_{0}^{2}+k^{2}\right)  \right]  \left(  \partial_{0}^{2}+k^{2}\right)
\delta R_{k}=8\pi G\delta\rho_{k}. \label{eq:Eq03}%
\end{equation}
This equation is a nonhomogeneous linear ordinary differential equation, and the stability analysis can
be studied by looking only at the homogeneous solution. So ignoring
$\delta\rho_{k}$, we have%
\begin{equation}
\frac{3}{\kappa_{0}}\left[  1+\frac{\beta_{0}}{\kappa_{0}}\left(  \partial
_{0}^{2}+k^{2}\right)  \right]  \left(  \partial_{0}^{2}+k^{2}\right)  \delta
R_{k}+\delta R_{k}=0. \label{Eq estabilidade base}%
\end{equation}
For $\beta_{0}=0$, we have%
\begin{equation}
\delta\ddot{R}_{k}+\omega_{k}^{2}\delta R_{k}=0,
\label{Estabilidade beta zero}%
\end{equation}
with%
\begin{equation}
\omega_{k}^{2}=k^{2}+\frac{\kappa_{0}}{3}.
\end{equation}
For $\kappa_{0}>0$, the perturbative solution\emph{ }$\delta R_{k}$\emph{ }is
oscillatory for all the\emph{ }$k$\emph{ }models and therefore, stable. On the
other hand, for $\kappa_{0}<0$ there are unstable $k$ modes that grow
exponentially, producing instabilities in the system.

Let us now look at the case of the complete equation. By manipulating Eq.
(\ref{Eq estabilidade base}), we can get%
\begin{equation}
\delta\ddddot{R}_{k}+\left(  \frac{\kappa_{0}}{\beta_{0}}+2k^{2}\right)
\delta\ddot{R}_{k}+\left(  k^{4}+\frac{\kappa_{0}}{\beta_{0}}k^{2}%
+\frac{\kappa_{0}^{2}}{3\beta_{0}}\right)  \delta R_{k}=0.
\end{equation}
By considering a solution for $\delta R_{k}$ in the form $e^{i\omega t}$, we
have%
\begin{equation}
\left[  \omega^{4}-\left(  \frac{\kappa_{0}}{\beta_{0}}+2k^{2}\right)
\omega^{2}+\left(  k^{4}+\frac{\kappa_{0}}{\beta_{0}}k^{2}+\frac{\kappa
	_{0}^{2}}{3\beta_{0}}\right)  \right]  \delta R_{k}=0,
\end{equation}
which results in%
\begin{equation}
\sqrt{2}\omega=\pm\sqrt{\left(  \frac{\kappa_{0}}{\beta_{0}}+2k^{2}\right)
	\pm\sqrt{\left(  \frac{\kappa_{0}}{\beta_{0}}+2k^{2}\right)  ^{2}-4\left(
		k^{4}+\frac{\kappa_{0}}{\beta_{0}}k^{2}+\frac{\kappa_{0}^{2}}{3\beta_{0}%
		}\right)  }}.
\end{equation}
The stability of the solutions occurs only if the four $\omega$'s were real
for any value of $k$. A necessary, but not sufficient,\emph{ }condition for
this to happen is%
\begin{equation}
\frac{\kappa_{0}}{\beta_{0}}+2k^{2}>0,
\end{equation}
which implies that\emph{ }$\kappa_{0}$\emph{ }and\emph{ }$\beta_{0}$\emph{
}must\emph{ }have the same sign. If this condition is met, we should still
have%
\[
0<4\left(  k^{4}+\frac{\kappa_{0}}{\beta_{0}}k^{2}+\frac{\kappa_{0}^{2}%
}{3\beta_{0}}\right)  <\left(  \frac{\kappa_{0}}{\beta_{0}}+2k^{2}\right)
^{2}.
\]
For $\beta_{0}$\emph{ }and\emph{ }$\kappa_{0}$\emph{ }both negative, the first
part of the inequality is not satisfied for all $k$. In fact, with $\beta
_{0}<0$, the expression%
\[
0<4\left(  k^{4}+\frac{\kappa_{0}}{\beta_{0}}k^{2}+\frac{\kappa_{0}^{2}%
}{3\beta_{0}}\right)  ,
\]
will be violated with a sufficiently small value of $k$. For $\beta_{0}$\emph{
}and\emph{ }$\kappa_{0}$\emph{ }both positive, the first part of the
inequality is always satisfied because all the terms of $k^{4}+\left(
\kappa_{0}/\beta_{0}\right)  k^{2}+\kappa_{0}^{2}/3\beta_{0}$ are positive.
Furthermore%
\begin{equation}
4\left(  k^{4}+\frac{\kappa_{0}}{\beta_{0}}k^{2}+\frac{\kappa_{0}^{2}}%
{3\beta_{0}}\right)  <\left(  \frac{\kappa_{0}}{\beta_{0}}+2k^{2}\right)
^{2}\Rightarrow\beta_{0}<\frac{3}{4}.
\end{equation}
Therefore, the stability of the solutions in the weak-field regime only occurs
if%
\begin{equation}
\kappa_{0}>0\text{ \ and \ }0\leq\beta_{0}<\frac{3}{4}. \label{eq:restriction}
\end{equation}

This analysis excludes oscillatory solutions since it restricts the range of
the parameters in $\kappa_{0}>0$ \ and \ $0\leq\beta_{0}<3/4$. As seen in the
previous section, in this range $l_{\pm}\in%
\mathbb{R}
$ and the corrections to the gravitational potential are Yukawa-type.

It is interesting to note that this last result can also be obtained by
considering a decomposition of the metric in the scalar and massless tensor modes. In
order to show that statement, we use some results found in Ref. \cite{Accioly:2016qeb}. In this case, the metric can be decomposed as
\begin{equation}
h_{\mu \nu }=\tilde{h}_{\mu \nu }-\eta _{\mu \nu }\Phi -\eta _{\mu \nu }\bar{%
	\Phi},  \label{eq:decomposicao}
\end{equation}%
assuming the harmonic gauge%
\begin{equation}
\partial ^{\mu }\tilde{\gamma}_{\mu \nu }=0\text{, \ \ with \ \  }\tilde{\gamma%
}_{\mu \nu }\equiv \tilde{h}_{\mu \nu }-\frac{1}{2}\eta _{\mu \nu }\tilde{h},
\end{equation}%
where $\tilde{h}=\eta ^{\alpha \beta }\tilde{h}_{\alpha \beta }$.
Furthermore, considering the correspondence between the parameters of the
theories and the difference of signature of the metric, we obtain the
equations%
\begin{gather}
\square \tilde{h}_{\mu \nu }=0, \\
\left( \square -m_{+}^{2}\right) \Phi =0, \\
\left( \square -m_{-}^{2}\right) \bar{\Phi}=-m_{+}^{2}\Phi ,
\end{gather}%
where for the scalar equations%
\begin{equation}
m_{\pm }^{2}=\frac{\kappa _{0}}{2\beta _{0}}\left( 1\pm \sqrt{1-\frac{4\beta
		_{0}}{3}}\right) .  \label{eq:m2pm}
\end{equation}%
For we have $m_{\pm }^{2}>0$, we conclude that $%
\kappa _{0}$ and $\beta _{0}$ must be restricted in the intervals expressed
in relation (\ref{eq:restriction}). Finally, note that the conditions that lead to $m_{\pm
}^{2}>0$, i.e., the absence of tachyons are the same that guarantee the stability
of the weak-field regime solutions.


\section{Black hole solutions \label{m:black-hole-solutions}}

In recent years, there has been growing interest in investigating black hole
solutions in higher-order theories of gravity \cite{Nelson:2010ig,Lu:2015cqa,PhysRevD.92.124019,Hennigar:2016gkm,Bueno:2016lrh,Goldstein:2017rxn,Bueno:2017sui,Hennigar:2017ego,Ahmed:2017jod,Bueno:2017qce,Kokkotas:2017zwt,Podolsky:2018pfe,Podolsky:2019gro,Hernandez-Lorenzo:2020aie}. Such
analyzes allow us to better understand, in addition to the structure and
nature of the solutions, these theories in strong gravitational field regimes.
We know that the spherically symmetric and static solution in a vacuum of GR
is the Schwarzschild one. Furthermore, Birkhoof's theorem tells us that it is
the only one. In this section, we investigate the existence of exterior
non-Schwarzschild black hole solutions, and we discuss under what conditions they may be
physically achievable.

By performing this analysis, we use an approach
called the Lichnerovicz method in the literature, similar to what is done in
Refs. \cite{Nelson:2010ig,Lu:2015cqa,PhysRevD.92.124019}. In a
nutshell, this methodology informs us about the possibility of the exterior
Schwarzschild solution to be unique --- which is interesting in our case,
since $R_{\mu\nu}=0$, that is, the Schwarzschild solution, is a solution of
field equations (\ref{eq:eqs-de-campo}) in a vacuum.

We start this investigation, considering the line element due to a
spherically symmetric and static object, which is written as%
\begin{equation}
ds^{2}=g_{00}dt^{2}+g_{11}dr^{2}+r^{2}\left(  d\theta^{2}+\sin^{2}\theta
d\phi^{2}\right)  ,
\end{equation}
where $g_{00}$ and $g_{11}$ are functions only of the radial
coordinate. Then let us take the trace of the field equations
(\ref{eq:eqs-de-campo}) in a vacuum, multiply it by the curvature scalar $R$
and use the Leibniz rule a few times to obtain total derivative terms. That
gives us%

\begin{widetext}
\begin{gather}
R^{2}+\frac{3}{\kappa_{0}}\nabla_{\mu}R\nabla^{\mu}R-\frac{3\beta_{0}}%
{2\kappa_{0}^{2}}R\nabla_{\mu}R\nabla^{\mu}R+\frac{3\beta_{0}}{\kappa_{0}^{2}%
}\square R\square R-\frac{\alpha_{0}}{3\kappa_{0}^{2}}R^{4}+\frac{3\alpha_{0}%
}{\kappa_{0}^{2}}\nabla_{\mu}R\nabla^{\mu}R^{2}+\nonumber\\
-\frac{3}{\kappa_{0}}\nabla_{\mu}\left(  R\nabla^{\mu}R\right)  +\frac
{3\beta_{0}}{\kappa_{0}^{2}}\nabla_{\mu}\left(  R\nabla^{\mu}\square R\right)
-\frac{3\beta_{0}}{\kappa_{0}^{2}}\nabla_{\mu}\left(  \nabla^{\mu}R\square
R\right)  +\frac{\beta_{0}}{\kappa_{0}^{2}}\nabla_{\mu}\left(  R^{2}%
\nabla^{\mu}R\right)  -\frac{3\alpha_{0}}{\kappa_{0}^{2}}\nabla_{\mu}\left(
R\nabla^{\mu}R^{2}\right)  =0.\label{eq:trace-eq-times-r}%
\end{gather}
\end{widetext}

Now, integrating Eq. (\ref{eq:trace-eq-times-r}) into the outside region of
the horizon $r_{h}$, where\emph{ }$g_{00}<0$ and\emph{ }$g_{11}>0$, we obtain%

\begin{widetext}
\begin{gather}
\int d^{4}x\sqrt{-g}\left[  R^{2}+\frac{3g^{11}}{\kappa_{0}}\left(
1-\frac{\beta_{0}}{2\kappa_{0}}R+\frac{2\alpha_{0}}{\kappa_{0}}R\right)
\left(  \partial_{1}R\right)  ^{2}+\frac{3\beta_{0}}{\kappa_{0}^{2}}\left(
\square R\right)  ^{2}-\frac{\alpha_{0}}{3\kappa_{0}^{2}}R^{4}\right]
+\nonumber\\
-\frac{3}{\kappa_{0}}K\left(  \int_{r_{h}}+\int_{\infty}\right)  dS_{1}\left(
r^{2}\sqrt{-g_{00}g^{11}}R\partial_{1}R\right)  +\frac{3\beta_{0}}{\kappa
	_{0}^{2}}K\left(  \int_{r_{h}}+\int_{\infty}\right)  dS_{1}\left[  r^{2}%
\sqrt{-g_{00}g^{11}}R\partial_{1}\left(  \square R\right)  \right]
+\nonumber\\
-\frac{3\beta_{0}}{\kappa_{0}^{2}}K\left(  \int_{r_{h}}+\int_{\infty}\right)
dS_{1}\left(  r^{2}\sqrt{-g_{00}g^{11}}\partial_{1}R\square R\right)
+\nonumber\\
+\frac{\beta_{0}}{\kappa_{0}^{2}}K\left(  \int_{r_{h}}+\int_{\infty}\right)
dS_{1}\left(  r^{2}\sqrt{-g_{00}g^{11}}R^{2}\partial_{1}R\right)
-\frac{3\alpha_{0}}{\kappa_{0}^{2}}K\left(  \int_{r_{h}}+\int_{\infty}\right)
dS_{1}\left(  r^{2}\sqrt{-g_{00}g^{11}}g^{11}R\partial_{1}R^{2}\right)=0,
\end{gather}
\end{widetext}
where the chosen notation in the second, third and fourth lines means that the
integration is evaluated at the spherically symmetric hypersurfaces $r=r_{h}$
and $r\rightarrow\infty$. Also, $K$ is a constant term involving angular and
temporal coordinates and%
\begin{equation}
\square R=\frac{1}{r^{2}\sqrt{-g_{00}g_{11}}}\partial_{1}\left(  r^{2}%
\sqrt{-g_{00}g_{11}}g^{11}\partial_{1}R\right)  .
\end{equation}
If we consider that the curvature scalar and its derivatives up to the
third-order are well-behaved quantities at $r=r_{h}$, which is reasonable
since the event horizon $r_{h}$ is an apparent singularity, we have that the
associated surface terms are null. This occurs because, by definition, an
event horizon is a null hypersurface in which $g^{11}\left(  r_{h}\right)
=g_{00}\left(  r_{h}\right)  =0$ \cite{Bekenstein:1971hc}. In turn, the
surface terms evaluated at infinity also vanish, since in this region far from
the origin, the expressions for the weak-field regime are valid, so that the
stable Yukawa-type solutions decay faster than any powers in $r$. Thus, we get%
\begin{widetext}
\begin{equation}
\int d^{4}x\sqrt{-g}\left[  R^{2}+\frac{3g^{11}}{\kappa_{0}}\left(
1-\frac{\beta_{0}}{2\kappa_{0}}R+\frac{2\alpha_{0}}{\kappa_{0}}R\right)
\left(  \partial_{1}R\right)  ^{2}+\frac{3\beta_{0}}{\kappa_{0}^{2}}\left(
\square R\right)  ^{2}-\frac{\alpha_{0}}{3\kappa_{0}^{2}}R^{4}\right]
=0.\label{eq:eq-base}%
\end{equation}
\end{widetext}
First,\emph{ }note that by making $\alpha_{0}=\beta_{0}=0$, which represents
the gravity with quadratic correction in the\emph{ }curvature scalar, we have%
\begin{equation}
\int d^{4}x\sqrt{-g}\left[  R^{2}+\frac{3g^{11}}{\kappa_{0}}\left(
\partial_{1}R\right)  ^{2}\right]  =0.\label{eq:resultado-lu-stelle}%
\end{equation}
Since $\kappa_{0}$, as well as $g^{11}$, is a positive quantity and $R^{2}$
and $\left(  \partial_{1}R\right)  ^{2}$ are quadratic terms, we note that to
satisfy the relation (\ref{eq:resultado-lu-stelle}), both terms must vanish
independently, in such a way that it gives $R=0$. By substituting this result
in Eq. (\ref{eq:eqs-de-campo}) free of source (with $\alpha_{0}=\beta_{0}=0$),
we find that $R_{\mu\nu}=0$. Thus, the Schwarzschild outer solution is the
unique spherically symmetric and static solution of the gravity with quadratic
correction in the\emph{ }curvature scalar.

Similarly, by following the same approach for the full relation
(\ref{eq:eq-base}), we observe initially that the first and third terms are
non-negative, since we saw in the last section that $\beta_{0}>0$. So, for the only spherically symmetric solution of the field equations
(\ref{eq:eqs-de-campo}) in vacuum to be the Schwarzschild solution, it is necessary that $\alpha_{0}\leq0$ and that the\emph{
}curvature scalar\emph{ }outside the horizon $R_{out}$ satisfies%
\begin{equation}
R_{out}\leq\frac{2\kappa_{0}}{4\left\vert \alpha_{0}\right\vert +\beta_{0}}.
\label{eq:desigualdade-R}%
\end{equation}
If these two conditions are fulfilled, all terms in Eq. (\ref{eq:eq-base})
will be non-negative, and to satisfy it, each term must vanish independently.
Then, we conclude that $R_{out}=0$. Based on (\ref{eq:desigualdade-R}), we can
state that if $R_{out}$ is less than or equal to such a quantity, then it will
necessarily be zero in that region. Unfortunately, since we cannot guarantee
that $R_{out}=0$ unrestrictedly, the Lichnerovicz method is quite inconclusive
in discriminating if the Schwarzschild solution is the only one.

On the other hand, we can also state that if $R_{out}$ exceeds this amount,
non-Schwarzschild black hole solutions might exist. Then let us estimate the
order of magnitude of $R_{out}$ for this to occur. The first consideration to
be made is that $\left(  4\left\vert \alpha_{0}\right\vert +\beta_{0}\right)
\sim\beta_{0}$ in (\ref{eq:desigualdade-R}). This is justified since $R^{3}$
and\emph{ }$R\square R$ are second-order correction terms, and therefore,
$\alpha_{0}$ and $\beta_{0}$ are expected to have similar magnitudes. Besides
that, the stability analysis of the weak-field limit solutions shows us that
there is an upper limit for the parameter $\beta_{0}$, namely the value of
$3/4$, which leads to a stable gravitational potential. At the limit where
$\beta_{0}$ assumes such a value, it follows that%
\begin{equation}
R_{out}\gtrsim\frac{8\kappa_{0}}{3}.
\end{equation}
If we assume that the additional terms in besides to the Einstein-Hilbert in the
action (\ref{eq:acao-da-teoria}) are classical corrections from a quantum
theory of gravity, it is reasonable to expect that $\kappa_{0}\sim M_{P}^{2}$.
Thus,%
\begin{equation}
R_{out}\gtrsim\frac{8\kappa_{0}}{3}\sim10^{36}%
\operatorname{GeV}%
^{2}.
\end{equation}

Another possibility is suppose that cosmic inflation is generated by the
modified gravity (\ref{eq:acao-da-teoria}). In this case, we can establish the
order of magnitude of $R_{out}$ based on the energy scale of inflation. By
using the results presented in Ref. \cite{Cuzinatto:2018vjt}, namely,%
\begin{equation}
\kappa_{0}\approx6\pi^{2}A_{s}rM_{P}^{2}\sim10^{-9}M_{P}^{2}\text{
	\ and\emph{\ \ }}\beta_{0}^{\max}\sim10^{-2}, \label{kappa0}%
\end{equation}
where the scalar amplitude\emph{ }$A_{s}\approx1.96\times10^{-9}$\emph{
}\cite{Akrami:2018odb} and the tensor-to-scalar ratio\emph{ }$r\sim10^{-2}%
$\emph{ }\cite{Cuzinatto:2018vjt}, we were able to estimate that%
\begin{equation}
R_{out}\gtrsim\frac{2\kappa_{0}}{\beta_{0}^{\max}}\sim\frac{10^{27}}{10^{-2}}=10^{29}%
\operatorname{GeV}%
^{2}. \label{Rout}%
\end{equation}
Thus, non-Schwarzschild black hole solutions may exist only for values of
$R_{out}$ which exceed\emph{ }$10^{29}%
\operatorname{GeV}%
^{2}$. Very roughly, we can estimate the order of magnitude of the horizon
extrapolating the validity of the weak-field results. Using Eqs.
(\ref{eq:proposta-curvatura-escalar}), (\ref{eq:b0}), (\ref{kappa0}) and
imposing the relation (\ref{Rout}), we obtain%
\begin{equation}
\frac{GM}{r}\left(  e^{-\frac{r}{l_{+}}}-e^{-\frac{r}{l_{-}}}\right)
\gtrsim10^{2}, \label{eq:eq2-analise-R}%
\end{equation}
where $l_{\pm}$ is given by Eq. (\ref{eq:raiz-li}). So, due to the exponential
decay, the necessary but not sufficient condition for\emph{ }$R_{out}%
\gtrsim10^{29}%
\operatorname{GeV}%
^{2}$\emph{ }is%
\begin{equation}
r\lesssim l_{\pm}\sim\sqrt{\frac{3}{2\kappa_{0}}}\Rightarrow r\lesssim
10^{4}M_{P}^{-1}\sim10^{-31}%
\operatorname{m}%
\text{.} \label{r estimation}%
\end{equation}
As it is expected that $R_{out}$ reaches its maximum value near to the
horizon, we get\emph{ }%
\[
\emph{\ }R_{out}\gtrsim10^{29}%
\operatorname{GeV}%
^{2}\Rightarrow r_{h}\lesssim10^{-31}%
\operatorname{m}%
.
\]
The previous analysis is quite rough and the estimate of $r_{h}$ can vary a
few orders of magnitude. Nevertheless, even considering this variation, it is
clear that the existence of non-Schwarzschild solutions will not occur in the
usual astrophysical context. At least for the values of $\kappa_{0}$ considered.

\subsection{Deviations from the Schwarzschild solution \label{m:deviations}}

We saw in the previous section that the use of the Lichnerovicz method proved
inconclusive in giving us an answer to the question of whether or not there
are exterior non-Schwarzschild solutions. We have also seen, by making use of
some estimations and extrapolations, that such solutions may be manifested
only in black holes whose horizons are extremely small. We now want to
investigate whether there are spherically symmetric and static solutions in
the vacuum which can be obtained from a continuous deformation of the
Schwarzschild solution, similar to that developed in Ref.
\cite{PhysRevD.92.124019}.

First of all, let's see how the trace of the field equations
(\ref{eq:eqs-de-campo}) in the vacuum behaves for the case where the curvature
scalar $R$ deviates infinitesimally from the Schwarzschild solution i.e.,%
\begin{equation}
R\left(  r\right)  =\varepsilon f\left(  r\right)  ,
\end{equation}
with $\varepsilon\ll1$. Discarding terms beyond the first-order in
$\varepsilon$, we have%
\begin{equation}
-f+\frac{3}{\kappa_{0}}\square_{Sch}f-3\frac{\beta_{0}}{\kappa_{0}^{2}}%
\square_{Sch}^{2}f=0,\label{eq:eq-perturbacao-ordem1-1}%
\end{equation}
where $\square_{Sch}$ is constructed with the Schwarzschild metric. Writing
explicitly the last two terms in Eq. (\ref{eq:eq-perturbacao-ordem1-1}), we
obtain
\begin{gather}
-f+\left\{  \frac{3}{\kappa_{0}}\frac{1}{r^{2}}\left[  2r\left(  1-\frac
{r_{h}}{r}\right)  +r_{h}\right]  +\frac{12\beta_{0}}{\kappa_{0}^{2}}%
\frac{r_{h}}{r^{4}}\left(  1-\frac{r_{h}}{r}\right)  \right\}  \frac{df}%
{dr}+\nonumber\\
+\left\{  \frac{3}{\kappa_{0}}\left(  1-\frac{r_{h}}{r}\right)  -\frac
{6\beta_{0}}{\kappa_{0}^{2}}\frac{1}{r^{2}}\left[  1-\left(  1-\frac{r_{h}}%
{r}\right)  ^{2}\right]  \right\}  \frac{d^{2}f}{dr^{2}}+\nonumber\\
-\frac{6\beta_{0}}{\kappa_{0}^{2}}\left[  \frac{1}{r}\left(  1-\frac{r_{h}}%
{r}\right)  ^{2}+\frac{1}{r}\left(  1-\frac{r_{h}}{r}\right)  +\frac{r_{h}%
}{r^{2}}\left(  1-\frac{r_{h}}{r}\right)  \right]  \frac{d^{3}f}{dr^{3}%
}+\nonumber\\
-\frac{3\beta_{0}}{\kappa_{0}^{2}}\left(  1-\frac{r_{h}}{r}\right)  ^{2}%
\frac{d^{4}f}{dr^{4}}=0.\label{eq:limite-lin-geral}%
\end{gather}
From the Eq. (\ref{eq:limite-lin-geral}), we make a construction similar to
that one developed in the previous section based on the Lichnerovicz method.
The idea is to set up a quadratic relation for $f$, integrate it outside the
horizon $r_{h}$, and check if we can conclude, through the independently vanishing of each of the terms, that $f$ is null in that region. In this
sense, by multiplying Eq. (\ref{eq:limite-lin-geral}) by $r^{5}$, we can
express it in the form%
\begin{equation}
h_{0}f+h_{1}f^{\prime}+h_{2}f^{\prime\prime}+h_{3}f^{\left(  3\right)  }%
+h_{4}f^{\left(  4\right)  }=0,\label{eq:limite-lin-geral-f}%
\end{equation}
where%
\begin{gather*}
h_{0}=-r^{5},\\
h_{1}=\frac{3}{\kappa_{0}}r^{3}\left(  2r-r_{h}\right)  +\frac{12\beta_{0}%
}{\kappa_{0}^{2}}r_{h}\left(  r-r_{h}\right)  ,\\
h_{2}=\frac{3}{\kappa_{0}}r^{4}\left(  r-r_{h}\right)  -\frac{6\beta_{0}%
}{\kappa_{0}^{2}}r\left[  r^{2}-\left(  r-r_{h}\right)  ^{2}\right]  ,\\
h_{3}=-\frac{12\beta_{0}}{\kappa_{0}^{2}}r^{3}\left(  r-r_{h}\right)  ,\\
h_{4}=-\frac{3\beta_{0}}{\kappa_{0}^{2}}r^{3}\left(  r-r_{h}\right)  ^{2}.
\end{gather*}
By multiplying Eq. (\ref{eq:limite-lin-geral-f}) by $u\left(  r\right)  f$,
where $u$ is a function whose form will be obtained later, we have%
\begin{equation}
h_{0}uf^{2}+h_{1}uff^{\prime}+h_{2}uff^{\prime\prime}+h_{3}uff^{\left(
	3\right)  }+h_{4}uff^{\left(  4\right)  }=0.\label{eq:limite-lin-geral-f-u}%
\end{equation}
Now, performing a series of derivative manipulations in Eq.
(\ref{eq:limite-lin-geral-f-u}), it is possible to rewrite it in terms of the
square of the functions $f$, $f^{\prime}$, $f^{\prime\prime}$, and total
derivative terms, as follows%
\begin{gather}
h_{0}uf^{2}+\left[  -h_{2}u+\frac{3}{2}\left(  h_{3}u\right)  ^{\prime
}-2\left(  h_{4}u\right)  ^{\prime\prime}\right]  f^{\prime2}+h_{4}%
uf^{\prime\prime2}+\nonumber\\
+\left[  h_{1}u-\left(  h_{2}u\right)  ^{\prime}+\left(  h_{3}u\right)
^{\prime\prime}-\left(  h_{4}u\right)  ^{\left(  3\right)  }\right]
ff^{\prime}+\nonumber\\
+\left(  h_{2}uff^{\prime}\right)  ^{\prime}+\left(  h_{3}uff^{\prime\prime
}\right)  ^{\prime}-\frac{1}{2}\left(  h_{3}uf^{\prime2}\right)  ^{\prime
}+\nonumber\\
-\left(  \left(  h_{3}u\right)  ^{\prime}ff^{\prime}\right)  ^{\prime}-\left(
h_{4}uf^{\prime}f^{\prime\prime}\right)  ^{\prime}+\left(  h_{4}uff^{\left(
	3\right)  }\right)  ^{\prime}+\nonumber\\
-\left(  \left(  h_{4}u\right)  ^{\prime}ff^{\prime\prime}\right)  ^{\prime
}+\left(  \left(  h_{4}u\right)  ^{\prime}f^{\prime2}\right)  ^{\prime
}+\left(  \left(  h_{4}u\right)  ^{\prime\prime}ff^{\prime}\right)  ^{\prime
}=0.\label{eq:limite-lin-geral-f-u-quad}%
\end{gather}
Then, we choose the form of $u\left(  r\right)  $ so that the quantity in
brackets accompanying the cross term $ff^{\prime}$ is canceled, which is,%
\begin{equation}
h_{1}u-\left(  h_{2}u\right)  ^{\prime}+\left(  h_{3}u\right)  ^{\prime\prime
}-\left(  h_{4}u\right)  ^{\left(  3\right)  }=0\Rightarrow u\propto1/r^{3}.
\end{equation}
By integrating Eq. (\ref{eq:limite-lin-geral-f-u-quad}) in the region outside
the horizon, we obtain that%
\begin{gather}
\int\left\{  r^{2}f^{2}+\left[  \frac{3}{\kappa_{0}}r\left(  r-r_{h}\right)
+\frac{6\beta_{0}}{\kappa_{0}^{2}}\frac{\left(  r-r_{h}\right)  ^{2}}{r^{2}%
}\right]  f^{\prime2}\right\}  dr+\nonumber \\ 
+\frac{3\beta_{0}}{\kappa_{0}^{2}}\int\left(  r-r_{h}\right)  ^{2}%
f^{\prime\prime2}dr+I_{S}=0, \label{eq:eq-base-desvio-linearizado1}%
\end{gather}
where $I_{S}$ represents the integral of all surface terms%
\begin{gather}
	I_{S}=\int_{r_{h}}^{\infty}\left[  \left(  \frac{h_{2}}{r^{3}}ff^{\prime
	}\right)  ^{\prime}+\left(  \frac{h_{3}}{r^{3}}ff^{\prime\prime}\right)
	^{\prime}-\frac{1}{2}\left(  \frac{h_{3}}{r^{3}}f^{\prime2}\right)  ^{\prime
	}\right]  dr+\nonumber\\
	-\int_{r_{h}}^{\infty}\left[  \left(  \left(  \frac{h_{3}}{r^{3}}\right)
	^{\prime}ff^{\prime}\right)  ^{\prime}+\left(  \frac{h_{4}}{r^{3}}f^{\prime
	}f^{\prime\prime}\right)  ^{\prime}\right]  dr+\nonumber\\
	+\int_{r_{h}}^{\infty}\left[  \left(  \frac{h_{4}}{r^{3}}ff^{\left(  3\right)
	}\right)  ^{\prime}-\left(  \left(  \frac{h_{4}}{r^{3}}\right)  ^{\prime
	}ff^{\prime\prime}\right)  ^{\prime}\right]  dr+\nonumber\\
	+\int_{r_{h}}^{\infty}\left[  \left(  \left(  \frac{h_{4}}{r^{3}}\right)
	^{\prime}f^{\prime2}\right)  ^{\prime}+\left(  \left(  \frac{h_{4}}{r^{3}%
	}\right)  ^{\prime\prime}ff^{\prime}\right)  ^{\prime}\right]  dr.
\end{gather}
The integration of the various surface terms evaluated at infinity vanishes
out, since in that region $f$ and its first derivatives have a Yukawa-like
behavior, decaying faster than any polynomial term. In turn, in order to
show that the surface terms cancel each other on the horizon, we only need to
consider that the function $f$ and its first derivatives are well behaved at
$r=r_{h}$. With that in mind, it is easy to show that at the horizon,%
\begin{gather}
I=-\left(  -\frac{6\beta_{0}}{\kappa_{0}^{2}}ff^{\prime}\right)  -\left(
\left(  0\right)  ff^{\prime\prime}\right)  +\frac{1}{2}\left(  \left(
0\right)  f^{\prime2}\right)  +\nonumber\\
+\left(  -\frac{12\beta_{0}}{\kappa_{0}^{2}}ff^{\prime}\right)  +\left(
\left(  0\right)  f^{\prime}f^{\prime\prime}\right)  -\left(  \left(
0\right)  ff^{\left(  3\right)  }\right)  +\nonumber\\
+\left(  \left(  0\right)  ff^{\prime\prime}\right)  -\left(  \left(
0\right)  f^{\prime2}\right)  -\left(  -\frac{6\beta_{0}}{\kappa_{0}^{2}%
}ff^{\prime}\right)  ,
\end{gather}
and therefore%
\begin{equation}
I=0.\nonumber
\end{equation}
Let us turn our attention back to Eq. (\ref{eq:eq-base-desvio-linearizado1}).
We observe that outside the horizon ($r>r_{h}$) each of the coefficients that
accompany the quadratic terms are non-negative. We see then that the only way
for Eq. (\ref{eq:eq-base-desvio-linearizado1}) to be satisfied is if each of
the terms is null independently, which is, if $f=f_{out}=0$. This is an
interesting result because it means that an infinitesimal perturbation of the
Schwarzschild exterior solution does not exist, at least not in the first-order.

The next step is to analyse higher-order perturbations of the Schwarzschild
solution. For this purpose, we can represent a perturbation from the Schwarzschild solution by%
\begin{equation}
g_{\mu\nu}=g_{\mu\nu}^{\left(  0\right)  }+\varepsilon g_{\mu\nu}^{\left(
	1\right)  }+\varepsilon^{2}g_{\mu\nu}^{\left(  2\right)  }+\cdots,
\label{eq:perturbacao-metrica}%
\end{equation}
where $\varepsilon$ is the parameter of the perturbation~and $g_{\mu\nu
}^{\left(  0\right)  }$ is the Schwarzschild metric. Such perturbation
represented in Eq. (\ref{eq:perturbacao-metrica}) induces a perturbation in
the curvature scalar $R$ given by%
\begin{equation}
R=R^{\left(  0\right)  }+\varepsilon R^{\left(  1\right)  }+\varepsilon
^{2}R^{\left(  2\right)  }+\cdots, \label{eq:perturbacao-curvatura-escalar}%
\end{equation}
in which $R^{\left(  0\right)  }=R_{Sch}=0$, and from the previous analysis,
$R_{out}^{\left(  1\right)  }\equiv f_{out}=0$. Taking the trace of the field
equation (\ref{eq:eqs-de-campo}) in the vacuum and considering terms up to the
second-order of perturbation, we obtain%
\begin{equation}
-R_{out}^{\left(  2\right)  }+\frac{3}{\kappa_{0}}\square_{Sch}R_{out}%
^{\left(  2\right)  }-3\frac{\beta_{0}}{\kappa_{0}^{2}}\square_{Sch}%
^{2}R_{out}^{\left(  2\right)  }=0. \label{eq:eq-perturbacao-ordem2-2}%
\end{equation}
The relation (\ref{eq:eq-perturbacao-ordem2-2}) is the same for the
first-order perturbation $f$, namely (\ref{eq:eq-perturbacao-ordem1-1}), whose
result we already know: it vanishes outside the horizon. Then we can conclude
that there is no $R_{out}^{\left(  2\right)  }$ perturbation. Extrapolating
this result it is possible to show that higher-order perturbations
$R_{out}^{\left(  n\right)  }$ are all null.

Based on the previous analysis, we could think that the result $R_{out}%
^{\left(  n\right)  }=0$ is valid for any $r_{h}$, but this is not true. The
main point is that the perturbative approach developed here configure a
regular perturbation analysis, and formally, this kind of analysis cannot be
employed in our case because the perturbative terms are differential
high-order terms. In this situation, the correct approach is to use singular
perturbation analysis (see Refs. \cite{book:5118,book:910642} for an introduction).
However, it is safe to employ usual regular perturbation if the region of
interest is far from the boundary layer region. For our model, the boundary
layer is located near to the origin and its thickness can be estimated as
$\kappa_{0}^{-1/2}$. As we are interested in outside horizon solutions, our
result, based on regular perturbative method, remain valid for $r_{h}\gg
\kappa_{0}^{-1/2}$. Therefore, we conclude that there are no deviations from
the Schwarzschild solution outside the horizon, since $r_{h}\gg
\kappa_{0}^{-1/2}$. This result is in agreement to the conclusion presented in
the previous section.

\section{Final comments \label{m:weak-field-limit copy(1)}
	\label{m:discussions}}

In this paper we have studied spherically symmetric and static solutions in
the weak-field regime and black hole context of a gravity theory which
includes all the curvature scalar terms up to the second-order corrections.

In the weak-field regime, we developed a complete study for the proposed
gravity, in which we obtained all possible solutions due to a point mass.
In addition, we verified that among all found solutions only those which present a
Yukawa-type behaviour ($\kappa_{0}>0$ \ and \ $0\leq\beta_{0}<3/4$) are stable
when a temporal perturbation is performed. It is worth mentioning, that the
same restriction for $\kappa_{0}$ and $\beta_{0}$ emerges if we consider that
the inflationary regime is generated by the action (\ref{eq:acao-da-teoria})
with $\alpha_{0}=0$ \cite{Cuzinatto:2018vjt}. In fact, the constraints
$\kappa_{0}>0$ \ and \ $0\leq\beta_{0}<3/4$ are necessary conditions to ensure
that the inflation properly end in a graceful exit. The consistency between
these two results is not completely unexpected as both address stability
issues. Nevertheless, their coherence provides an important clue in
constraining effective theories of quantum gravity.

Motivated by the growing interest in black hole solutions in higher-order
gravities, see Refs. \cite{Nelson:2010ig,Lu:2015cqa,PhysRevD.92.124019,Hennigar:2016gkm,Bueno:2016lrh,Goldstein:2017rxn,Bueno:2017sui,Hennigar:2017ego,Ahmed:2017jod,Bueno:2017qce,Kokkotas:2017zwt,Podolsky:2018pfe,Podolsky:2019gro,Hernandez-Lorenzo:2020aie}, we also developed an analysis
of spherically symmetric and static black holes solutions. In this
investigation, we used the Lichnerovicz method to investigate the possibility
of existence of exterior non-Schwarzschild solutions. In both approaches used, namely, 1) considering the trace equation of the full theory and 2) assuming a continuous deformation from the exterior Schwarzschild solution, our results suggest the absence of non-Schwarzschild macroscopical black holes. This is due to the fact that the curvature scalar $R_{out}$ must exceed, in the most conservative estimate, $10^{29}%
\operatorname{GeV}%
^{2}$ for the existence of non-Schwarzschild solutions, and this implies in black holes whose horizons are smaller or of the order of $10^{-31}%
\operatorname{m}%
$.
In this context, there are two main questions which should be addressed. First, do these mini black holes actually exist, since $R_{out} > 10^{29} 
\operatorname{GeV}%
^{2}$ is a necessary but not sufficient condition for its existence? Second, how do they behave under Hawking radiation (see Ref. \cite{Konoplya:2019ppy} and references therein), and consequently, what would be the lifetime of these objects? These two issues will be addressed by the authors in a future work.

\section*{Acknowledgments}
G. Rodrigues-da-Silva thanks CAPES/UFRN-RN (Brazil) for financial support and L. G. Medeiros acknowledge CNPq (Brazil) for partial financial support.

\bibliography{references}
\end{document}